\documentclass[12pt]{article}
\usepackage{times}
\usepackage{amsmath}
\usepackage{amsfonts}
\usepackage{mathrsfs}
\usepackage{array}
\usepackage{amssymb}
\usepackage{graphicx}
\begin{document}
\pagestyle{plain}
\hsize = 6.5 in
\vsize = 8.5 in
\hoffset = -0.75 in
\voffset = -0.5 in
\baselineskip = 0.29 in
\def\rd{{\rm d}}

\def\vb {{\bf b}}
\def\vj {{\bf j}}
\def\vP{{\bf p}}
\def\vp{{\bf p}}
\def\vx{{\bf x}}
\def\vu{{\bf u}}
\def\vv{{\bf v}}
\def\vw {{\bf w}}
\def\mA{{\bf A}}
\def\mB{{\bf B}}
\def\mC{{\bf C}}
\def\mD{{\bf D}}
\def\mG{{\bf G}}
\def\mI{{\bf I}}
\def\vJ{{\bf J}}
\def\mM{{\bf M}}
\def\mS{{\bf S}}
\def\mQ{{\bf Q}}
\def\mR{{\bf R}}
\def\mT{{\bf T}}
\def\mU{{\bf U}}
\def\mV{{\bf V}}
\def\mGa{\mbox{\boldmath$\Gamma$}}
\def\mPhi{\mbox{\boldmath$\Phi$}}
\def\mPi{\mbox{\boldmath$\Pi$}}
\def\mXi{\mbox{\boldmath$\Xi$}}
\def\vg {\mbox{\boldmath$\gamma$}}
\def\vpi{\mbox{\boldmath$\pi$}}
\renewcommand{\thefootnote}{\fnsymbol{footnote}}
\def\vF {{\bf F}}
\def\vg {\mbox{\boldmath$\gamma$}}
\def\vA {{\bf A}}
\def\vB {{\bf B}}
\def\vD {{\bf D}}

\title{Thermodynamics of the General Diffusion
Process: Equilibrium Supercurrent and Nonequilibrium Driven
Circulation with Dissipation}
\author{Hong Qian\\[9pt]
Department of Applied Mathematics\\
University of Washington\\
Seattle, WA 98195-3925, USA
}

\maketitle


\begin{abstract}

\end{abstract}
Unbalanced probability circulation, which yields cyclic motions in phase space, 
is the defining characteristics of a stationary diffusion process without detailed balance.  In over-damped soft matter systems, such behavior is a hallmark of 
the presence of a sustained external driving force accompanied with dissipations.  In an under-damped and strongly correlated system, however,
cyclic motions are often the consequences of a conservative dynamics.   
In the present paper, we give a novel interpretation of a class of diffusion
processes with stationary circulation in terms of a Maxwell-Boltzmann equilibrium 
in which cyclic motions are on the level set of stationary probability density 
function thus non-dissipative, e.g., a supercurrent.  This implies an 
orthogonality between stationary circulation $J^{ss}(x)$ and the gradient of stationary probability density $f^{ss}(x)>0$.   A sufficient and necessary
condition for the orthogonality is a decomposition of the drift 
$b(x)=j(x)+\mD(x)\nabla\varphi(x)$ where $\nabla\cdot j(x)=0$
and $j(x)\cdot\nabla\varphi(x)=0$.  Stationary processes with 
such Maxwell-Boltzmann equilibrium has an underlying 
conservative dynamics 
$\dot{x}= j(x)\equiv$ $\big(f^{ss}(x)\big)^{-1}J^{ss}(x)$, 
and a first integral $\varphi(x)\equiv-\ln f^{ss}(x)=$ const, 
akin to a Hamiltonian system.  At all time, an instantaneous free 
energy balance equation exists for a given diffusion system; and 
an extended energy conservation law among an entire 
family of diffusion 
processes with different parameter $\alpha$ can be established 
via a Helmholtz theorem.  For the general diffusion process
without the orthogonality, a nonequilibrium cycle 
emerges, which consists of external driven $\varphi$-ascending 
steps and spontaneous $\varphi$-descending movements, alternated 
with iso-$\varphi$ motions.   The theory presented here 
provides a rich mathematical narrative for complex mesoscopic 
dynamics, with contradistinction to an earlier one [H. Qian et. al., {\em J. Stat. Phys.} {\bf 107}, 1129 (2002)].

\section{Introduction}

P. W. Anderson, J. J. Hopfield, and many other condensed matter 
physicists have all pointed out that emergent phenomena at each and 
every different scales actually obey different laws
which require research that is just as fundamental in its nature 
as any other research \cite{anderson_72,jjh_94,laughlin_00}.
The intricate behavior of a complex dynamic is particularly pronounced
at a mesoscopic scale which contains too many individual ``bodies''
from a Newton-Laplacian perspective but not sufficient many, and often 
too strong an interaction and too heterogeneous, for the universal 
statistical laws such as central limit theorems and Gaussian 
processes to apply. 

	This paper provides a didactic mathematical narrative 
of the general diffusion process, as a concrete model for 
complex stochastic nonlinear dynamics, in the light of two 
very different thermodynamic interpretations.
The first one has its root in over-damped soft matters where 
sustained cyclic motions are considered a driven phenomenon 
accompanied with dissipation.  The second one is motivated 
by under-damped and strongly correlated systems in 
which oscillatory motions are often the consequences of
a conservative dynamics.   

Throughout the paper, classical thermodynamic terminologies 
are introduced with precise mathematical definitions in the 
framework of the general diffusion process.  They are 
perfectly consistent with equilibrium statistical mechanics 
\cite{rtcox_1,lebowitz} and mostly in accord with known notions 
in nonequilibrium statistical physics. Their ultimate validity, 
of course, are judged by the internal logic of the 
mathematics.  Indeed, there is a growing awareness that,
in order to fully develop a thermodynamic theory for 
mesoscopic nonequilibrium systems, its foundation has to be 
shifted away from the empirical notion of local equilibrium first 
formulated by the Brussel school, toward a mathematical 
theory.  In the present work, it is the Markov dynamics \cite{qqt,qian_jmp,qian_pla}. 

	As expressed by some high-energy physicists ``the rest is 
chemistry'' \cite{anderson_72}; 
one can indeed learn from chemistry a very powerful 
perspective on complex systems around us:  
First,  classical chemistry distinguish itself from 
physics by quantifying a dynamical system in terms of ``species'' ,
``individuals'', and the numbers of individuals in a particular species,
rather than tracking the detailed particle positions and velocities.  
This practice is consistent with many-body physics in
which Eulerian rather than Lagrangian description of a fluid, and 
second quantization, are prefered.  This was a fundamental insight 
of Boltzmann who, together with Maxwell, Gibbs, Smoluchowski, 
Einstein, and Langevin, paved the way to use stochastic 
mathematics as the proper language for quantifying complex systems.

	Second, while nonequilibrium thermodynamics in homogeneous
systems usually deals with temperature and pressure gradients, nonequilibrium chemical thermodynamics often deals with 
isothermal, isobaric systems with
all kinds of interesting phenomena, including animated living matters,
under chemical potential differences.  Chemical equilibrium is actually 
an isothermal, dynamic concept.

	Third, perhaps the most profound insight from chemistry, is the
recognition of emerging discrete states, and transitions among them
in molecules | each one a nonlinear continuous many-atom system
in its own right: 
Such a state is sufficiently stable against small perturbations of the 
underlying equations of atomic motion to be identified as a 
distinct ``chemical species''.  Such a transition is on an entirely 
different time scale; it necessarily crosses a barrier 
chemists called a ``transition state''.  The rare event can be 
quantified in terms of an exponentially distributed random time 
and the notion of a ``reaction
coordinate'' or an ``order parameter''.  This is a great achievement
in multi-scale modeling by separation of time scales.

	Finally, and possibly a deep idea from chemistry, is that
stationary probability, as an emergent statistical entity, can actually
be formulated as a law of force that quantifies collective motion 
of a system.  Entropic force arises from mere probabilistic 
descriptions; and the concept of ``potential of mean force'' first
articulated by J. G. Kirkwood in the theory of fluid mixtures 
\cite{kirkwood_35} is the ultimate explanation of 
equilibrium free energy; and it is actually a conditional probability!
Chemical potential difference can do mechanical work; it can
power a ``Maxwell demon'' \cite{tuyh}.  

	While the chemistry providing a perspective, 
the mathematical theory of stochastic processes fully 
developed in the first half of 20th century provides a 
powerful analytical tool for representing complex dynamics. 
In fact, many key notions in chemistry echo important concepts 
in the theory of probability \cite{qian_kou}.  In recent years,
a nonequilibrium steady state with positive entropy production that is 
consistent with over-damped soft matters and biochemical systems 
\cite{NP_book,Hill_book} has been mathematically defined in terms 
of Markov processes \cite{zqq,gqq,jqq}.  Stochastic thermodynamics 
has emerged as a unifying theory of nonequilibrium statistical 
mechanics \cite{esposito,gehao,seifert,jarzynski}.  Classical phase transition 
theory can be understood using elementary chemical master
equations and stochastic differential equations
with bistability in the limit of both time and system's 
size tending to infinities \cite{aqtw,ge_qian_prl}.

Indeed, stochastic dynamics, which formalizes rapid stochastic and
slower nonlinear dynamics and privides dual descriptions on both 
individual trajectories and ensemble probability distributions, 
seems to be a natural match for Anderson's 
hierarchical structure of sciencs generated by
symmetry breakings \cite{aqtw,qian_qb_13}.   Even
the Universe has become only one of the individuals of
a multiverse \cite{briangreene_book}.

This paper is structured as follows:
In Sec. \ref{sec2}, results following the first perspective
are summarized \cite{qqt,qian_jmp,jqq}.  
This is the main story line of the theory of 
stochastic thermodynamics which goes much further to fully
explore trajectory-based thermodynamics and fluctuation theorems 
\cite{esposito,gehao,seifert,jarzynski}.  Sec. \ref{sec3} begins 
with introducing the notion of Maxwell-Boltzmann (MB)
equilibrium with non-dissipative supercurrent, and presents the 
defining characteristics of diffusion processes that possess an
MB equilibrium with circulations: (i) an orthogonality between
stationary current $J^{ss}$ and the gradient of the stationary potential
$\varphi(x)=-\ln f^{ss}(x)$; (ii) a decomposition of 
$b(x)=j(x)-\mD(x)\nabla\varphi(x)$ where $\nabla\cdot j(x)=0$
and $j(x)\cdot\nabla\varphi(x)=0$.
Sec. \ref{sec4} investigates the emergent divergence-free vector 
field $j(x,\alpha)$ from a family of diffusion processes with MB 
equilibrium: $\dot{x}=j(x,\alpha)$ has a first integral $\varphi(x,\alpha)$.
The Helmholtz theorem is applied to establish an
extended energy conservation law $h=h(\sigma_B,\alpha)$ 
for the entire family of diffusion processes, among the 
$\varphi$-level sets of which $h$ is the energy and
$\sigma_B$ is the Boltzmann entropy.
Sec. 5 studies diffusion processes that do not meet
the orthogonality condition.  We argue that the stationary
process of such a system has both external driving
force and dissipation, thus it is a nonequilibrium steady
state within the framework of under-damped thermodynamics.
Sec. 6 provides some discussions.

\section{Diffusion processes with  and without 
circulation}
\label{sec2}

We conresider a family of stochastic, diffusion process $X_{\alpha\beta}(t)$ 
with a transition probability density function $f_{\alpha\beta}(x,t|y)$
that satisfies
Fokker-Planck equation
\begin{equation}
     \frac{\partial f_{\alpha\beta}(x,t)}{\partial t} = \nabla\cdot\Big(
                         \beta^{-1}\mD(x,\alpha)\nabla f_{\alpha\beta}(x,t) - b(x,\alpha) f_{\alpha\beta}(x,t) \Big),
\label{the_eq}
\end{equation}
with non-local boundary condition and initial data
\begin{equation}
     \int_{\mathbb{R}^n} f_{\alpha\beta}(x,t)\rd x = 1; \    \
                f_{\alpha\beta}(x,0) = \delta(x-y),
\label{ibc}
\end{equation}
in which $x, y, b\in\mathbb{R}^n$, and $\mD$ is a $n\times n$
positive definite matrix.  The $\alpha$ is a continuous parameter
that defines the family of related diffusion processes; and the $\beta$ 
is a scaling parameter quantifying the magnitude of the ``noise''. 
We shall assume that a positive, steady state probability density 
function exists
\begin{equation}
     \lim_{t\rightarrow\infty} f_{\alpha\beta}(x,t|y) = f_{\alpha\beta}^{ss}(x),
\end{equation}
which is independent of $y$ and statisfies the 
stationary Fokker-Planck equation
\begin{equation}
        \beta^{-1}\mD(x,\alpha)\nabla f_{\alpha\beta}^{ss}(x) 
       - b(x,\alpha) f_{\alpha\beta}^{ss}(x) \equiv -J^{ss}(x), \    \
              \nabla\cdot J^{ss}(x) = 0,
\label{eq_4}
\end{equation}
under the same boundary condition in (\ref{ibc}).
For more discussions on the mathematical setup of this
problem, see \cite{qqt,jqq}.

	The following facts are known under 
appropriate mathematical conditions.  All discussions in
Sec. \ref{sec2} and Sec. \ref{sec3} assume a fixed value 
of $\alpha$, which we shall suppress until Sec. \ref{sec4}.

\subsection{Diffusion processes with detailed balance}
\label{sec2.1}

 The system (\ref{the_eq}) is called 
detailed balanced if $J^{ss}(x)=0$ $\forall x\in\mathbb{R}^n$.  
This is true if and only if a $\varphi(x)$ exists such that
$\mD^{-1}(x)b(x)=-\nabla
\varphi(x)$.  Then $f^{ss}(x)=Z^{-1}(\beta) e^{-\beta\varphi(x)}$
where $\varphi(x)$ is a potential energy of the system, and 
$Z(\beta)$ is a normalization factor:
\begin{equation}
      Z(\beta) =  \int_{\mathbb{R}^n} e^{-\beta\varphi(x)}\rd x.
\end{equation}
Then the quantity
\begin{eqnarray}
         F(t) &=& \frac{1}{\beta}
                \int_{\mathbb{R}^n} f(x,t)\ln\left(\frac{f(x,t)}{f^{ss}(x)}\right)
                  \rd x
\label{eq6}\\
	&=&  \big\langle \varphi(x) \big\rangle - \beta^{-1}
                \left(-\int_{\mathbb{R}^n} f(x,t)\ln f(x,t)\rd x\right)
           +\beta^{-1}\ln Z(\beta),
\label{eq7}
\end{eqnarray}
in which we introduced the notion $\langle\cdots\rangle$ 
as the expected value with respect to time-dependent
probability distribution $f(x,t)$.  The first term in (\ref{eq7}) is 
the mean energy of the system at time $t$, and the term in the
parenthesis is Gibbs-Shannon entropy.  Therefore, it is
natural to indentify the $F(t)$ as an instantaneous,
generalized free energy of the dynamical system at time $t$.   
Actually, $F(t)$ is defined with respect to the equilibrium free 
energy of the system:  $F(t)\ge 0$ and it is zero when 
$f(x,t)=f^{ss}(x)$.  Classical statistical mechanics of inanimate 
matters uses universal mechanical energy as a reference point; 
thus the equilibrium free energy is 
$-\beta^{-1}\ln Z(\beta)$.\footnote{In classical statistical mechanics,
Newtonian mechanical energy is given {\em a priori}.  Then 
the potential condition $\mD^{-1}(x)b(x)=-\nabla\varphi(x)$ becomes the
fluctuation-dissipation relation.  It is an essential equation 
completing a phenomenological theory of equilibrium fluctuations
in terms of a diffusion process.  It has the same nature as the
detailed balance condition in discrete-state Markov process 
models widely used in chemistry \cite{lewis25}. 
}

	As a function of time, it can be mathematically shown that 
\begin{equation}
    \frac{\rd F(t)}{\rd t} = \int_{\mathbb{R}^n} J(x,t) 
                \nabla\mu(x,t) \le 0,
\label{dFdt}
\end{equation}
in which $\mu(x,t)=\varphi(x)+\beta^{-1}\ln f(x,t)$,
$J(x,t)= -f(x,t)\mD(x)\nabla\mu(x,t)$.  $\mu(x,t)$
can and should be interpreted as a generalized chemical potential,
or thermodynamic force, and $J(x,t)$ as the corresponding
thermodynamic flux.  $F(t)$ monotonically decreases
until reaching its minimum zero.   
In fact, $e_p(t)\equiv -\beta\frac{\rd F(t)}{\rd t} \ge 0$
is called entropy production rate for the diffusion 
process \cite{jqq,qqt}.

	One also has
\begin{equation}
    \frac{\rd S(t)}{\rd t} \equiv 
             \frac{\rd}{\rd t}\left(-\int_{\mathbb{R}^n} f(x,t)\ln f(x,t)\rd x\right)
     = e_p(t) + \beta\frac{\rd}{\rd t}  \big\langle \varphi(x) \big\rangle,
\label{eq9}
\end{equation}
the right-hand-side of which are entropy production, usually 
written as $\frac{\rd_iS}{\rd t}$ which is not a total differential, 
and $\frac{\rd_e S}{\rd t}$ is the heat flux due to exchange with
the environment \cite{NP_book}.   Nonequilibrium entropy balance
equation like (\ref{eq9}) was first put forward phenomenologically 
by the Belgian thermodynamist de Donder, founder of the 
Brussels School  \cite{coveney,tolman}.  The shift from using 
state function entropy to free energy, and the fact that  
$e_p(t)\equiv -\beta\frac{\rd F(t)}{\rd t}$, reflect Helmhotz's contribution to the Second Law as ``free energy decreases'' for a
canonical system, not ``entropy increases''; but  the origin of 
decreasing free energy is still the same positive entropy production.  

	The mathematical theory of stochastic processes offers
additional insights for systems with detailed balance:
A stationary stochastic trajectory $X(t)$ is time-reversible in 
a statistical sense \cite{qqt,jqq}.  Therefore, anything 
accomplished through a sequence of 
events has an equal probability of being undone; nothing
can be accomplished in an equilibrium dynamics.  The
linear operator on the right-hand-side of (\ref{the_eq}) is
self-adjoint; there can be no oscillatory 
dynamics, only multi-exponential decays.

\subsection{Diffusion processes with unbalanced circulation}

	A diffusion process without detailed balance has
$J^{ss}(x)\neq 0$, but $\nabla\cdot J^{ss}(x)=0$.  The
system has unbalanced probability circulation in 
the stationary state as its hallmark.  The two mathematical
objects: $\varphi_{\beta}(x)\equiv -\beta^{-1}\ln f^{ss}_{\beta}(x)$
and $J^{ss}_{\beta}(x)$ can be understood in analogous to 
the potential and current in an electrical system.  Note that
$\varphi_{\beta}(x)$ is now also a function of $\beta$, and
its limit when $\beta\rightarrow\infty$ can be highly non-smooth.
Still, if one introduces $F(t)$ as in (\ref{eq6}), then 
Eq. (\ref{dFdt}) becomes
\begin{equation}
  \frac{\rd F(t)}{\rd t} = E_{in}(t) - e_p(t),
\label{eq10}
\end{equation}
where
\begin{eqnarray}
   E_{in}(t) &=& \int_{\mathbb{R}^n} J(x,t)
                      \Big( \mD^{-1}(x) b(x)- \beta^{-1}\nabla\ln f^{ss}_{\beta}(x)\Big) \rd x,
\label{11}\\
	e_p(t) &=&  \int_{\mathbb{R}^n} J(x,t)
                          \Big( \mD^{-1}(x) b(x)- \beta^{-1}\nabla\ln f(x,t)\Big) \rd x.
\label{12}
\end{eqnarray}
All three quantities have definitive sign \cite{esposito_07,gehao_09,
ge_qian_10,esposito_vandenbroeck_10}:
\begin{equation}
     \frac{\rd F(t)}{\rd t} \le 0,  \   \
      E_{in}(t) \ge 0,  \   \  
      e_p(t) \ge 0.
\label{eq13}
\end{equation}
Detailed balance  holds if and only if $E_{in}(t)=0$, stationarity 
holds if and only if $\frac{\rd F(t)}{\rd t}=0$, and $e_p(t)=0$ implies
both.  There are two verbal interpretations for Eqs. \ref{eq10}--\ref{eq13}:
(\ref{eq10}) can be read as a generalized nonequilibrium free energy 
balance equation with instantaneous energy source from its 
environment $E_{in}(t)$ and dissipation $e_p(t)$ \cite{qian_jmp,qian_pla}.
Alternatively, $e_p(t) = -\frac{\rd F(t)}{\rd t} + E_{in}(t)$ 
can be read as total entropy production has two distinct origins: 
the spontaneous self-organization into stationary state and 
the continuous environmental drive that keeps the system away
from its equilibrium.  The two terms correspond nicely to Boltzmann's
thesis and Prigogine's thesis on irreversibility, respectively.
Quasi-steady state is a conceptual device that bridges these two 
views \cite{ge_qian_2013}.

\subsection{Diffusion operator decomposition}
\label{sec2.3}

	The right-hand-side of (\ref{the_eq}) is a second-order
linear differential operator,
\begin{equation}
	\mathscr{L} \big[u\big] =  \nabla\cdot\Big(
                         \beta^{-1}\mD(x,\alpha)\nabla u(x) - b(x,\alpha) u(x) \Big),
\end{equation}
in an appropriate Hilbert space $\mathscr{H}$, with inner product 
\begin{equation}
           \big\langle u,v\big\rangle = \int_{\mathbb{R}}
                      u(x)v(x)\big(f^{ss}(x)\big)^{-1} \rd x,  \   \
                      u,v\in\mathscr{H}.
\end{equation}
Then $\mathscr{L}$ is self-adjoint if and only if the 
diffusion process is detail balanced \cite{qqt,jqq}.    Furthermore
$\mathscr{L}=\mathscr{L}_S + \mathscr{L}_A$,
where 
\[
           \Big\langle \mathscr{L}_S [u],v\Big\rangle =
                    \Big\langle u,\mathscr{L}_S[v]\Big\rangle, \   \
		 \Big\langle \mathscr{L}_A [u],v\Big\rangle =
                    -\Big\langle u,\mathscr{L}_A[v]\Big\rangle,
\]
$\forall u,v\in\mathscr{H}$.  

A diffusion process with self-adjoint
$\mathscr{L}_S$ has $E_{in}(t)=0$.  

A degenerated diffusion
with skew symmetric $\mathscr{L}_A$ has $\frac{\rd F(t)}{\rd t}=0$
for all $t$.  It actually has a non-random dynamics  whose 
trajectories follow the ordinary differential equation
$\dot{x}=\big(f_{\beta}^{ss}(x)\big)^{-1}J_{\beta}^{ss}(x)$ \cite{qian_jmp}. 
The solution curves of this equation in phase space is identical to
$\dot{x}=J^{ss}_{\beta}(x)$, $\nabla\cdot J^{ss}_{\beta}(x)=0$.
Paradoxically, such a dynamical system is called ``conservative''
in classical mechanics.
 
	Because the diffusion with $\mathscr{L}_A$ is degenerate, 
both $E_{in}(t)$ and $e_p(t)$ in (\ref{11}) and (\ref{12}) are 
infinite thus no longer defined.

	The mathematics of decomposing $\mathscr{L}$ 
is not new {\em per se} \cite{vankampen,risken}, but its 
clear relation with nonequilibrium thermodynamics and the 
theory of entropy production is novel \cite{qian_jmp,qian_pla}.

\section{\boldmath{$\nabla\varphi\perp J^{ss}$}:
Circulation as a supercurrent in a Maxwell-Boltzmann equilibrium}
\label{sec3}

	All the mathematical narritive so far fits established
chemical thermodynamics of over-damped molecular systems.  
In particular, when applied to a molecular motor, the
$E_{in}$ term in (\ref{eq10}) is indeed the amount  of ATP
hydrolysis free energy, and $e_p$ the heat dissipation \cite{qian_jpcm}.

In physics, however, the notion of a persistent current describes
a perpetual electrical current without requiring an external 
power source.  A superconducting current is one example.
We now show that an alternative, novel thermodynamic 
interpretation based on an under-damped dynamic perspective 
is equally legitimate \cite{qian_pla}:  $J^{ss}(x)=0$ is no longer the 
defining characteristics of an equilibrium.  Rather we define a 
Maxwell-Boltzmann (MB) equilibrium as 
$J^{ss}(x)\cdot \nabla\varphi(x) = 0$ 
where $\varphi(x)=-\beta^{-1}\ln f_{\beta}^{ss}(x)$. 
Unbalanced circulation on an equal-$\varphi$ level set is 
considered conservative.
 
The following facts are known under 
appropriate mathematical conditions.

\subsection{$\varphi(x)$ is independent of $\beta$}

	We assume, when $\beta=1$, $J_{1}^{ss}(x)\perp\nabla\varphi_{1}(x)$.  
This implies also an orthogonality between 
$J_{1}^{ss}(x)$ and $\nabla f_{1}^{ss}(x)$
$\forall x\in\mathbb{R}^n$.  One can decompose vector field 
$b(x)$ as
\begin{equation}
        b(x)  =  \big( f^{ss}_1(x)\big)^{-1}J_{1}^{ss}(x) + 
            \mD(x)\nabla\ln f^{ss}_1(x),
\label{016} 
\end{equation}
which can be re-written as
\begin{equation}
    b(x)  =  \big( f^{ss}_1(x)\big)^{-\beta}
                \left(  \big( f^{ss}_1(x)\big)^{-1+\beta} J_{1}^{ss}(x) \right)
               + \beta^{-1}\mD(x)\nabla\ln \big( f^{ss}_1(x) \big)^{\beta}. 
\end{equation}
Since
\[
       \nabla\cdot   \left(  \big( f^{ss}_1(x)\big)^{-1+\beta} J_{1}^{ss}(x) \right)
           = 0,
\]
we identify 
$\big( f^{ss}_1(x)\big)^{-1+\beta} J_{1}^{ss}(x)=J^{ss}_{\beta}(x)$
and  $\big(f_1^{ss}(x)\big)^{\beta}=f_{\beta}^{ss}(x)$, which is a
solution to (\ref{eq_4}).   Therefore, 
$f^{ss}_{\beta}(x)=Z^{-1}(\beta)e^{-\beta\varphi(x)}$.  Furthermore, 
$J_{\beta}^{ss}(x)=j(x)e^{-\beta\varphi(x)}$
in which $j(x)$ is also independent of $\beta$ 
and divergence free:
\[
      \nabla\cdot j(x)=e^{\beta\varphi(x)}\nabla\cdot J^{ss}_{\beta}(x)
 + J^{ss}_{\beta}(x)\cdot\nabla e^{\beta\varphi(x)} = 0.
\]
The diffusion process in (\ref{the_eq}) has an MB equilibrium
if and only if \cite{qian_pla}
\begin{equation}
             b(x)=j(x)-\mD(x)\nabla\varphi(x), \   \
                  \nabla\cdot j(x) = 0, \   \
                 j(x)\cdot\nabla\varphi(x) = 0.
\label{the_cond}
\end{equation}
This result generalizes the potential condition in Sec. \ref{sec2.1}
with an additional divergence-free, orthogonal $j(x)$.  
(\ref{the_cond}) is a much more restrictive condition on $b(x)$
then $b(x) = \big(f^{ss}_{\beta}(x)\big)^{-1}J^{ss}_{\beta}(x)-\beta^{-1}\mD(x)\nabla\ln f_{\beta}^{ss}(x)$,
which is valid for any (\ref{the_eq}) with a stationary
$f_{\beta}^{ss}(x)$ \cite{wangjin_1,wangjin_2}.
In general, when $\beta\rightarrow\infty$, the existence and 
characterizations of the limits of 
$\mu_{\beta}(\omega)=\int_{\omega}f_{\beta}^{ss}(x)\rd x$ and 
$\varphi_{\beta}(x)=-\beta^{-1}\ln f_{\beta}^{ss}(x)$ are highly non-trivial.

\subsection{Nonlinear dynamics $j(x)$}

	The unbalanced stationary circulation in a 
MB equilibrium $J^{ss}_{\beta}(x)=j(x)e^{\beta\varphi(x)}$
has a clear deterministic, underlying nonlinear dynamics
\begin{equation}
              \frac{\rd x}{\rd t} = j(x),  \   \    \nabla\cdot j(x)=0.
\label{hdynamics}
\end{equation}
Zero divergence of the vector field $j(x)$ means the dynamics is 
volume preserving in phase space.  Furthermore, $\varphi(x)$
is one conserved quantity:
\begin{equation}
   \frac{\rd}{\rd t} \varphi\big(x(t)\big) = \nabla\varphi(x) 
               \cdot\left(\frac{\rd x}{dt}\right)
                 = \nabla\varphi(x) \cdot j(x) = 0.
\end{equation}
Therefore, the dynamics in Eq. \ref{hdynamics} is akin
to a Hamiltonian system.   There is an agreement between
the stochastic thermodynamics and the nonlinear
dynamics.  Indeed, the operator decomposition
in Sec. \ref{sec2.3}, $\mathscr{L}=\mathscr{L}_S+\mathscr{L}_A$
matches the vector field decomposition
$b(x)=j(x)-\mD(x)\nabla\varphi(x)$:
\begin{eqnarray}
     \mathscr{L}_S\big[u\big] &=& \nabla\cdot\Big[
                    \beta^{-1}\mD\nabla\ln 
                         \Big(u(x) e^{\beta\varphi(x)}\Big) u(x) \Big],
\\
	\mathscr{L}_A\big[u\big] &=& -\nabla\cdot\Big(j(x)u(x)\Big).
\end{eqnarray}

\subsection{Entropy production}
	
	It has been shown that for a diffusion processes with MB equilibrium,
the entropy production rate that is consistent with both known physics 
and the stochastic trajectory-based mathematical formulation 
based on time reversal is the free energy decreasing rate, 
or non-adiabatic entropy production \cite{qian_pla}:
\begin{eqnarray}
           \frac{\rd F(t)}{\rd t}  &=&  \int_{\mathbb{R}^n} J(x,t)
            \Big(\nabla\varphi(x)+\beta^{-1}\nabla\ln f(x,t) \Big) \rd x                  
\\
          &=& -\int_{\mathbb{R}^n}  \nabla\mu(x,t)
                  \mD(x)\nabla\mu(x,t) f(x,t)\rd x,
\label{epformula}
\end{eqnarray}
in which again $\mu(x,t)=\varphi(x)+\beta^{-1}\ln f(x,t)$, as 
in Eq. \ref{dFdt}.  We note that even though 
$J(x,t) = j(x)f(x,t)-f(x,t)\mD(x)\nabla\mu(x,t)$ contains the 
conservative current $j(x)$, it has completely disappeared 
in the final entropy production formula (\ref{epformula}).

	Secondly, the rate of mean energy change:
\begin{eqnarray}
	\frac{\rd}{\rd t}\big\langle\varphi (x) \big\rangle &=&
             \int_{\mathbb{R}^n} J(x,t) \nabla\varphi(x) \rd x                  
\\
          &=& \beta^{-2}\int_{\mathbb{R}^n}  \nabla \big(e^{\beta\mu(x,t)}\big)
                  \mD(x) \nabla \big(e^{-\beta\varphi(x)}\big) \rd x.
\end{eqnarray}

The meanings of the two quantities $E_{in}(t)$ and $e_p(t)$ 
are yet to be elucidated for the under-damped systems.  
The mathematical expression for stationary $E_{in}$ 
and $e_p$ in (\ref{11})  and (\ref{12}) now reads
\begin{equation}
       \int_{\mathbb{R}^n}  j(x) \mD^{-1}(x) j(x) e^{-\beta\varphi(x)}\rd x.
\end{equation}
It has a resemblance to kinetic energy; one chould argue that
in an under-damped thermal mechanical equilibrium, kinetic energy 
comes in and heat goes out.

\subsection{Three examples}

	We now give three examples of diffusion processes,
with increassing generality, that have an MB 
equilibrium \cite{qian_pla}.

{\bf\em  Ornstein-Uhlenbeck process.} 
As a Gaussian Markov process, the Ornstein-Uhlenbeck (OU) process 
is the most widely used stochastic-process model in science and 
engineering \cite{wax,rrfox_78}.  Interestingly, its stationary process is 
always an MB equilibrium.  Realizing that stationary OU process is 
the universal theory for linear stochastic dynamics 
\cite{rtcox_2,onsager_machlup,lax}, this result  will
have far-reaching implications. 

	An OU process has a constant diffusion matrix $\mD$ and 
a linear $b(x) = -\mB x$.
All the eigenvalues of $\mB$ are assumed to have positive real
parts.   The stationary probability density and circulation can be
exactly computed in terms of the covariant matrix $\mXi$
of a Gaussian distribution:  
\begin{subequations}
\begin{eqnarray}
              f^{ss}(x) &=& \big(2\pi \big)^{-\frac{n}{2}} \Big(\det(\mXi)\Big)^{-\frac{1}{2}}
                       \exp\left(-\frac{1}{2}x^T\mXi^{-1}x\right),
\\
	J^{ss}(x) &=& \big(\mB-\mD\mXi^{-1}\big)x f^{ss}(x), 
\\
               &&   \mB\mXi + \mXi \mB^T = 2\mD.
\end{eqnarray}
\label{OUP}
\end{subequations}
$J^{ss}(x)=0$ if and only if $\mB\mD=\mD\mB^T$ \cite{qian_prsa}.  Then $\mXi=\mB^{-1}\mD$.  In general, the solution to the Lyapunov matrix 
equation (\ref{OUP}c) has an integral representation
\[
         \mXi = 2\int_0^{\infty} e^{-\mB s}\mD e^{-\mB^T s} \rd s.
\]
Noting that $\mXi \mB^T-\mD = \mD-\mB\mXi$ is anti-symmetric \cite{kat},
\begin{eqnarray*}
           J^{ss}(x)\cdot\nabla f^{ss}(x) &=& \Big(
                 \big(\mB-\mD\mXi^{-1}\big)x f^{ss}(x)\Big)^T
                        \cdot \nabla f^{ss}(x)
\\
	&=& - \Big(f^{ss}(x) \Big)^2  \Big(x^T
             \big(\mB-\mD\mXi^{-1}\big)^T \mXi^{-1} x\Big)
\\
	&=& - \Big(f^{ss}(x) \Big)^2  x^T\mXi^{-1} \Big(
           \mXi\mB^T-\mD \Big)\mXi^{-1}x \ = \  0.
\end{eqnarray*}
In fact, the linear vector field $-\mB x$, which has a decomposition $\big(\mD\mXi^{-1}-\mB\big)x-\mD\mXi^{-1}x$ as in (\ref{the_cond}), 
can be further represented as \cite{kat}
\begin{equation}
             -\mB x =     -\big(\mA+\mD\big)\nabla\left(\frac{1}{2}x^T\mXi^{-1}x\right).
\label{022}
\end{equation}
in which matrix $\mA+\mD$ has a symmtric part $\mD$
and an anti-symmetric part $\mA=\mB\mXi-\mD$.  In other words, any linear
vector field $b(x)=-(\mA+\mD)\nabla\varphi(x)$ where $\varphi(x)$ is quadratic.

	The corresponding linear stochastic differential equation
\[
     \rd X(t) = -(\mA+\mD)\nabla\varphi(x)\rd t 
                   + \big(2\mD\big)^{\frac{1}{2}} \rd B(t),
\]
then, can be re-written as 
\begin{subequations}
\label{eqn25}
\begin{equation}
       \mM \rd X(t) = -\nabla\varphi(x) \rd t + \mGa\rd B(t), 
\end{equation}
in which $\mM=(\mA+\mD)^{-1}$, $\mGa=\mM(2\mD)^{\frac{1}{2}}$, 
and
\begin{equation}
       \mGa\mGa^T = 2\mM\mD\mM^T 
                              = \mM+\mM^T,
\end{equation}
which is twice the symmetric part of $\mM$.  Stochastic differential 
equations expressed such as (\ref{eqn25}), which describes
a stochastic dynamics without detailed balance but still 
reaching an MB equilibrium $e^{-\varphi(x)}$, first appeared 
in \cite{ao}.
\end{subequations}

{\bf\em Klein-Kramers equation.}
The Klein-Kramers equation is the canonical stochastic Newtonian
dynamics with a stochastic damping that satisfies fluctuation-dissipation 
relation, thus a Maxwell-Boltzmann distribution in its stationary state.  It 
has
\begin{equation}
          \mD = k_BT\left(\begin{array}{cc}  
                          0 & 0  \\   0  & \eta(x)  \end{array} \right), \  
            b(x,y) =  \left(\begin{array}{l}  
                          \  \   m^{-1}y    \\[5pt]  -U'(x)
                         -m^{-1}\eta(x) y    \end{array} \right),
\end{equation}
and $f^{ss}(x,y)=\exp\big(-\frac{H(x,y)}{k_BT}\big)$ where
the Hamiltonian function $H(x,y)$ is the total mechanical energy
\begin{equation}
    H(x,y) = \frac{y^2}{2m} + U(x),
\end{equation} 
and
\begin{equation}
          J^{ss}(x,y) =  \left(\begin{array}{c}
                           m^{-1}y \\[5pt]  -U'(x) \end{array}\right)f^{ss}(x,y),
  \    \
                 j(x,y) = \left(\begin{array}{c}
                          \partial H/\partial y \\[5pt]  -\partial H/\partial x \end{array}\right).  
\end{equation}
$\frac{\rd}{\rd t}(x,y)^T=j(x,y)$ is indeed the underlying Hamiltonian
dynamics.  The stationary stochastic circulation $J^{ss}(x)$ is the 
Hamiltonian conservative dynamics weighted by the
stationary probability $f^{ss}(x,y)$.

{\bf\em P. Ao's stochastic process.}   P. Ao and his coworkers have 
generalized the Eq. \ref{022} and the Kramers-Klein equation to 
a class of nonlinear diffusion processes \cite{kat,ao,yin_ao,ao_qian} with:
\begin{equation}
		\mD(x) = \frac{1}{2}
                        \Big(\mG(x)+\mG^T(x)\Big),  \  
	      b(x) =  -\mG(x)\nabla\varphi(x).
\end{equation}
The stationary probability density and circulation are again
readily obtained:
\begin{equation}
     f^{ss}(x) = e^{-\varphi(x)},   \ 
    J^{ss}(x) = \frac{1}{2}\Big(\mG(x)-\mG^T(x)\Big)\nabla f^{ss}(x).
\end{equation}
Obviously, $J^{ss}(x)\cdot\nabla f^{ss}(x)=0$.  Three
interesting mathematical questions arise from this 
model: 

(i)  For any divergence-free vector field $j(x)$ with a first integral 
$\varphi(x)$, e.g., $j(x)\perp\nabla\varphi(x)$,
whether there always exists an anti-symmetrix $\mA(x)$ such that
$j(x)=\mA(x)\nabla\varphi(x)$?   This is true for any Hamiltonian
system, with even $n$, where $\mA$ is  symplectic.  For
$n=3$, such a $j(x)$ is called a gradient conjugate 
system, whose solution curves are all closed orbits \cite{zhangjy}.

(ii) For a vector field $b(x)$ that satisfies the decomposition in 
(\ref{the_cond}), what is the relationship between a 
Sinai-Bowen-Ruelle type invariant measure, when it exists, 
and the the $e^{-\beta\varphi(x)}$ in the limit of 
$\beta^{-1}\rightarrow 0$  \cite{youngls}?

(iii) For any vector field $b(x)$, whether there always exists an 
symmetric matrix $\mD(x)$ such that 
$b(x)=j(x)-\mD(x)\nabla\varphi(x)$ with diverence-free $j(x)$ and
$j(x)\perp\nabla\varphi(x)$?

\section{Helmholtz theorem and Carnot cycle}
\label{sec4}

	We now consider the family of diffusion processes in terms of 
the parameter $\alpha$ in (\ref{the_eq}).  All the discussions in 
Sec. \ref{sec2} and Sec. \ref{sec3} have been focused on the 
dynamics and thermodynamics of one autonomous dynamical 
system, or time-homogeneous stochastic process, with a 
single $\alpha$ value.  Particularly, we have noticed
that the generalized, nonequilibrium free energy $F(t)$
in (\ref{eq10}) takes its stationary state as the reference
point.  Classical thermodynamics, however, is a
theory of relationships among different stationary
states connected through ``changing a parameter''.
Being able to provide a common energy reference point for
diffusion processes with different $\alpha$, one 
needs a unique $\varphi(x,\alpha)$ which can not be 
obtained unambiguously from the analysis presented 
so far in Sec. \ref{sec3}.

\subsection{Generalizing Helmholtz theorem}

	One should recognize that the long-time ``state'' 
of $x(t)$ following the equation of motion (\ref{hdynamics}) is not a 
single point in $\mathbb{R}^n$; but a bounded orbit confined 
in a particular level set of $\varphi$; it is actually ``a state
of stationary motion'' \cite{gallavotti_book}.   This distinguishes 
a microscopic state in classical mechanics from a macroscopic 
state in classical thermodynamics.  In general, $x(t)$ is not 
ergodic on the entire level set.  However, realizing
that $j(x)$ is only a deterministic representation of the
stochastic circulation, one expects a time-scale separation
between the intra-$\varphi$-level-set motion and motion
across level sets \cite{ma_qian}.  To represent and characterize
an entire $\varphi$ level set, Boltzmann's idea was to
quantify it using some geometric quantities.  The 
volume it contains in $\mathbb{R}^n$ is one example:
\begin{equation}
            \sigma_B(h,\alpha) = \ln \left( 
                        \int_{\varphi(x,\alpha)\le h}               		    
                      \rd x \right).
\label{031}
\end{equation}
An elementary calculation of probability yields
\begin{equation}
   -\beta^{-1}\ln\int_{h<\varphi(x,\alpha)\le h+\rd h} e^{-\beta\varphi(x,\alpha)}
             \rd x = \varphi(x,\alpha)
                         -\beta^{-1}\ln\int_{h<\varphi(x,\alpha)\le h+\rd h} \rd x.
\label{032}
\end{equation}
Since ``free energy is equal to internal energy minus 
temperature times entropy'', recognizing the last term in (\ref{032})
as entropy is almost mandatory.  Then,
\begin{eqnarray}
         Z_{\alpha}(\beta) &=& \int_{\mathbb{R}^n} e^{-\beta\varphi(x,\alpha)}
              \rd x  \ = \ \int_{-\infty}^{\infty}     e^{-\beta h+\sigma_B(h,\alpha)} 
		\left(\frac{\partial\sigma_B}{\partial h}\right)_{\alpha} \rd h
\label{033} \\
	&=&  \int_{-\infty}^{\infty}  e^{-\beta h} \left(
                           \oint_{\varphi(x,\alpha)=h} \frac{\rd S}{\|\nabla_x\varphi(x,\alpha\|}
                         \right) \rd h. 
\label{034}                         
\end{eqnarray}
The  term inside the parenthesis in (\ref{034})  is known as 
microcanonical partition function \cite{khinchin_book}.
Eq. \ref{034} relates the microcanonical partition function to 
the canonical partition function $Z_{\alpha}(\beta)$.

	We now apply the Helmholtz theorem \cite{gallavotti_book} 
for the family of conservative dynamics $j(x,\alpha)$ with parameter 
$\alpha$, which has a conserved quantity $\varphi(x,\alpha)$
\cite{ma_qian}.  Note that due to its definition, the vector 
field $j(x,\alpha)$ is structurally stable within the framework
of the general diffusion process (\ref{the_eq}).

	$\sigma_B$ in (\ref{031}) is an increasing 
function of $h$, thus implicit function theorem 
applies to $\sigma_B(h,\alpha)$.  Thus we have
$h=h(\sigma_B,\alpha)$, and 	
\begin{equation}
    \rd h = \left(\frac{\partial h}{\partial\sigma_B}\right)_{\alpha}\rd\sigma_B
              +\left(\frac{\partial h}{\partial\alpha}\right)_{\sigma_B}\rd\alpha,
\label{1stlaw}
\end{equation}
in which
\begin{eqnarray}
     \left(\frac{\partial\sigma_B}{\partial h}\right)_{\alpha}
          &=& \frac{\displaystyle \oint_{\varphi(x,\alpha)=h} 
                 \frac{ \rd S}{\|\nabla_x\varphi(x,\alpha)\|}
                     }{\displaystyle \int_{\varphi(x,\alpha)\le h} 
                  \rd x  } 
\label{eq_36}\\
            &=& \left[ \frac{\displaystyle \oint_{\varphi(x,\alpha)=h} 
                   x_k\left(\frac{\partial\varphi(x,\alpha)}{\partial x_k}
                   \right)_{x_{j,j\neq k},\alpha}\frac{ \rd S }{\|\nabla_x\varphi(x,\alpha)\|}   
                             }{\displaystyle \oint_{\varphi(x,\alpha)=h} 
                 \frac{ \rd S}{\|\nabla_x\varphi(x,\alpha)\|} } \right]^{-1},
\label{eq_37}
\end{eqnarray}
which is the same for all $k=1,2,\cdots,n$.  This is known as the virial 
theorem in classical mechanics.  Gauss-Ostrogradsky-Green's 
divergence theorem is used to derive (\ref{eq_37}) from (\ref{eq_36}).

 We shall denote
\begin{equation}
           \theta(h,\alpha) \equiv \left( \frac{ \partial\sigma_B}{\partial h} \right)_{\alpha}^{-1},
             \   \ 
          F_{\alpha}(h,\alpha) \equiv -\left(\frac{\partial h}{\partial\alpha}\right)_{\sigma_B}.
\label{theta_F}
\end{equation}
Then,
\begin{equation}
		   \frac{F_{\alpha}}{\theta}   
          =  -\left(\frac{\partial h}{\partial\alpha}\right)_{\sigma_B}
              \left(\frac{\partial\sigma_B}{\partial h}\right)_{\alpha}
             =
                \left(\frac{\partial\sigma_B}{\partial\alpha}\right)_h.
\end{equation}

	Historically, the significance of the Helmholtz theorem is 
generalizing mechanical energy conservation to the First Law of Thermodynamics.  It provided a mechanical theory of heat
\cite{gallavotti_book}.  In the present theory, our Eq. \ref{1stlaw} 
has extended conservative $\varphi(x,\alpha)$ defined for 
each stochastic dynamical system with a particular $\alpha$ to a 
much broader energy conservation law $h(\sigma_B,\alpha)$ 
among the entire family of dynamical systems with different $\alpha$.

	{\bf\em Free energy and entropy of a Gaussian
distribution.}
For a $n$-dimensional Gaussian distribution with covariance matrix 
$\mXi(\alpha)$, the quadratic potential function $\varphi(x)$  
has the form $\frac{1}{2}\sum_{k=1}^n \lambda_k^{-1}\xi_k^2$
under an orthonormal transformation, 
where $\lambda$s are the eigenvalues of $\mXi$.  Then
\begin{equation}
         \sigma_B(h)= \ln\left(V_n
               \prod_{k=1}^n\sqrt{2h\lambda_k} \right) =
                     \frac{n}{2} \ln h + \frac{1}{2}\ln\det\big(\mXi\big) 
                       +\frac{n}{2}\ln 2+ \ln V_n,
\label{4gaussian}
\end{equation}
in which $V_n= \pi^{n/2}\Gamma^{-1}\big(\frac{n}{2}+1\big)$ is the
volume of an $n$-dimensional Euclidean ball with radius $1$.   

	Free energy of the same quadratic $\varphi(x)$, according to
canonical partition function, is
\[
	 -\beta^{-1}\ln \int_{\mathbb{R}^n}  \exp\left(-\frac{\beta}{2}x^T\mXi^{-1} x\right) 
        \rd x =
    -\beta^{-1} \left\{
          \frac{n}{2}\ln \big(2\pi\beta^{-1}\big) + 
                 \frac{1}{2}\ln\det\big(\mXi\big)
                 \right\}.
\]
Mean internal energy is $\overline{h}=\frac{n}{2\beta}$, and entropy is
\[        
          \frac{n}{2}\ln \overline{h} + 
                 \frac{1}{2}\ln\det\big(\mXi\big) + \frac{n}{2}+\frac{n}{2}
                      \ln\left(\frac{4\pi}{n}\right),
\]
which agrees with the $\sigma_B(h,\alpha)$ in (\ref{4gaussian}) when
$n$ is large, via Stirling's formula.

If the determinant of $\mXi(\alpha)$ is linearly dependent upon a 
parameter $\alpha$, then, 
$(\frac{\partial\sigma_B}{\partial\alpha})_h=(2\alpha)^{-1}$, and
$\alpha F_{\alpha} = \frac{1}{2}\theta$.  One could identify $\theta$ 
with temperature, $\alpha$ as volume, and $F_{\alpha}$ as pressure, 
then this relation is the law of ideal gas in classical thermodynamics.
Eqs. \ref{theta_F} and \ref{4gaussian} also give $\theta=2h/n$.
$h$ being a function of $\theta$ alone is known as Joule's law, which
states that the internal energy of an ideal gas is a function only of its
temperature.

	Based on these observations, it is not unreasonable to 
suggest the OU process as a mesoscopic dynamic model for 
stochastic thermodynamic behavior of an ideal system, such as 
ideal gases and ideal solutions \cite{ma_qian_2}. 

{\bf\em \boldmath{$\beta$} as an ensemble average of
\boldmath{$\left\langle\left(\partial\sigma_B/\partial h\right)_{\alpha} \right\rangle$}. }
Note that the right-hand-side of (\ref{033}) can also be re-written into
a different expression:
\begin{equation}
   Z_{\alpha}(\beta) = \int_{-\infty}^{\infty}   e^{-\beta  h+\sigma_B(h,\alpha)}  
               \beta \rd h,
\end{equation}
in which $\beta$ plays the role of 
$\left( \frac{ \partial\sigma_B}{\partial h} \right)_{\alpha}$
in (\ref{033}).  $\beta$, in fact, can be expressed implicitly 
as an ensemble average of 
$\left( \frac{ \partial\sigma_B}{\partial h} \right)_{\alpha}$:
\begin{equation}
      \beta = \frac{1}{Z_{\alpha}(\beta)}
                   \int_{-\infty}^{\infty} e^{-\beta h+\sigma_B(h,\alpha)} 
		\left(\frac{\partial\sigma_B}{\partial h}\right)_{\alpha} \rd h.
\end{equation}

\subsection{Carnot cycle in a family of 
diffusion processes with MB equilibrium}

Functional relations among triple quantities such
as $\big(\sigma_B,h,\alpha\big)$,
$\big(\sigma_B,F_{\alpha},\alpha\big)$, and 
$\big(\theta,F_{\alpha},\alpha\big)$ are called 
``equations of state'' in the classical thermodynamics.
They are powerful mathematical tools quantifying
long-time behavior of a family of conservative dynamics
$j(x,\alpha)$.

A Carnot cycle consists of two iso-$\theta$ curves and
two iso-$\sigma_B$ curves in $\alpha$ versus $F_{\alpha}$ 
plane.

	Let us again consider the simple model 
$\sigma_B(h,\alpha)=\mu\ln h+\nu\ln\alpha$.  Then we have
equations for iso-$\theta$ curves and iso-$\sigma_B$ curves:
\begin{equation}
        \alpha F_{\alpha} = \nu\theta,   \textrm{ and }  \
        \alpha^{1+\nu/\mu} F_{\alpha} =   \frac{\nu}{\mu} e^{\sigma_B/\mu}.	
\end{equation}

\section{Externally driven cycle of a diffusion process with dissipative 
circulation}

	The foregoing discussion clearly points to 
two mathematical objects $J^{ss}(x)$ and 
$\varphi(x)=-\beta^{-1}\ln f^{ss}(x)$ associated with a stationary
diffusion process.   We shall set $\beta=1$ in the following
discussion.  The steamlines of vector
field $J^{ss}(x)$ and the level sets of $\varphi(x)$ are 
perfect matched in a Maxwell-Boltzmann equilibrium
with non-dissipative circulation.

\begin{figure}[h]
\vskip 0.7cm
\begin{center}
\includegraphics[width=5.25in]{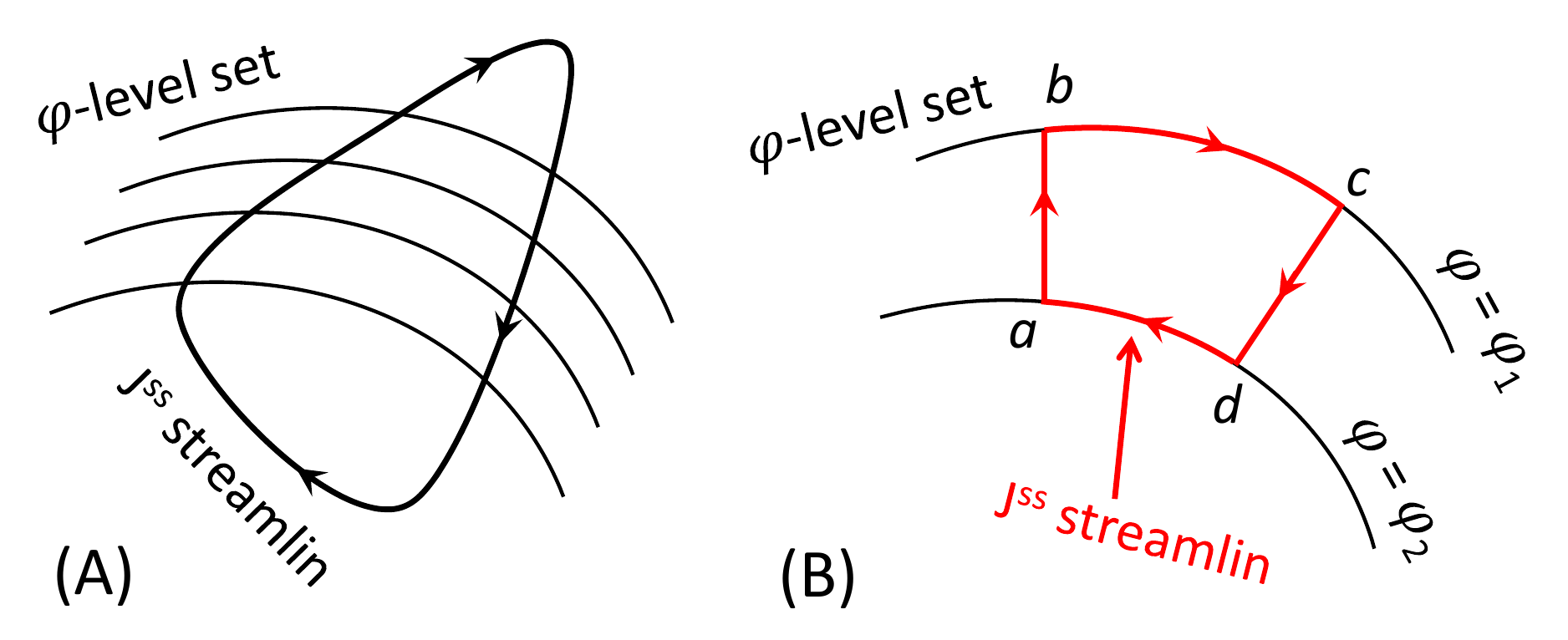}
\end{center}
\caption{(A)  Diffusion processes that have non-orthogonal
$J^{ss}(x)$ and $\nabla\varphi(x)$ have the streamlines of $J^{ss}$
passing through different level sets of $\varphi$.  Being a conservative
dynamics, the streamlines of vector field $J^{ss}$ have almost closed 
orbits according to Poincar\'{e} recurrence theorem.  
Any closed orbit can be approximated by portions that are
confined in $\varphi$-level sets, and portions that perpendicular
to $\varphi$-level sets.  (B)  An idealized red closed loop $abcda$  
consists four steps.  Assuming $\varphi_1>\varphi_2$,
then step $ab$ decreases in probability, thus it in general has to be
externally driven; steps $bc$ and $da$ are confined in $\varphi$-level
sets, thus they are conservative; step $cd$ increases in probability, 
thus it is spontaneous with dissipation.
}
\label{fig_1}
\end{figure}

	If the condition (\ref{the_cond}) is not met, then a diffusion 
process has stationary circulation with dissipation.  This implies
the process also has to be externally driven.  Thus, its 
stationary process is in a nonequilibrium steady state, as illustrated
in Fig. \ref{fig_1}(A).  See \cite{noh_lee} for a recent paper on
the subject.   One can in fact idealize any closed orbit
in phase space into four pieces as shown in Fig. \ref{fig_1}(B):
Movement from $c\rightarrow d$ is a spontaneous relaxation 
from high $\varphi$-level to low $\varphi$-level accompanied 
with dissipation; it is followed by a non-dissipative step 
from $d\rightarrow a$ confined in the level set $\varphi(x)=\varphi_2$;
then followed by a transition $a\rightarrow b$ from low $\varphi$-level 
to high $\varphi$-level, which has to driven by an external force;
and finally another non-dissipative move from $b\rightarrow c$ 
confined in the level set $\varphi(x)=\varphi_1$.   

	The analysis of energetic steps discussed above, and 
illustrated in Fig. \ref{fig_1}, is analogous to 
a pendulum system with damping and being driven:
\begin{equation}
      m\frac{\rd^2 x}{\rd t^2} = -k\sin x - \eta\left(\frac{\rd x}{\rd t}\right)
                + \xi(t),
\end{equation}
in which $\xi(t)$ is an oscillatory driving force.
In the absence of damping $-\eta\dot{x}$ and driving force
$\xi(t)$, the mechanical energy level sets are
\[
              H\big(x,\dot{x}\big) = \frac{m}{2}\dot{x}^2 + 
                                   k\Big(1-\cos x\Big).
\]
Then,
\begin{equation}
     \frac{\rd}{\rd t} H\big(x,\dot{x}\big) = -\eta\dot{x}^2 
                   + \dot{x}\xi(t).
\label{038}
\end{equation}
When the right-hand-side of (\ref{038}) is positive, 
the system gains energy; and when it is negative,
the system dissipates energy.  Over a complete cycle, 
these two terms have to balance with each other.  We
also noticed that they are even and odd
functions of $\dot{x}$, respectively.

	One could argue that the orthogonal relation
between $j(x)$ and $\nabla\varphi(x)$, for all $x$, in a system
with MB equilibrium is a ``local equilibrium condition''.
Such a condition provides a mathematical basis for
organizing the entire $\mathbb{R}^n$ state space, and the 
conservative motions, in terms of a single scalar $\varphi(x)$.   It
has been suggested as a possible mathematical statement of the
Zeroth Law of thermodynamics \cite{qian_pla}.  In 
contrast, the system in Fig. \ref{fig_1}
has its level sets and streamlines ``out of equilibrium''.

\section{Discussion}

	One interesting implication of the
present work, perhaps, is that a linear stochastic
dynamics is always consistent with a Maxwell-Boltzmann
equilibrium, together with a Hamiltonian system-like
conservative dynamics \cite{kat,qian_pla}.    This result 
completely unifies the stochatsic thermodynamic theory and 
the linear phenomenological approaches to 
equilibrium fluctuations pioneered by Einstein and
Onsager, and extended by many others, e.g., R. Kubo, 
Landau-Lifshitz, H. B. Callen, M. Lax, and J. Keizer, to name
a few.

	It is important to realize, therefore, that near a stable
dynamic fixed point, one can {\em not} determine the 
nature of equilibrium vs. nonequilibrium fluctuations in a 
subsystem from the internal data {\em alone}.   Additional information 
concerning the external environment is required to uniquely 
select one of the two possible {\em thermodynamics}.  In fact, 
the origin of the entropy production in an overdamped
nonequilibrium steady state is outside the subsystem, as
the notions of source and sink clearly imply.  Therefore, a 
nonequilibrium steady state of a subsystem can only be ``fully 
understood'' by including its envirnment; and for a universe 
without an outside, a underdamped thermodynamics 
with the heat death is the only logic consequence.

	We have recently suggested that the mathematical 
description of stochastic dynamics is an appropriate analytical
framework for P. W. Anderson's hierarchical structure of
science \cite{aqtw}.   The origin of randomness has been widely
discussed by many scholars; for example Poincar\'{e} has stated in 
1908 that  \cite{poincare_08} ``A very small cause, which 
escapes us, determines a considerable effect which we cannot 
ignore, and we then say that this effect is due to chance.'' 
In the light of \cite{anderson_72,jjh_94,laughlin_00}, this
might be updated: A very complex collection of causes, 
which are not understood by us, determines an un-avoidable 
consequence which we cannot ignore, and we then say that this 
effect is due  to chance, or our ignorance.

The mathematical approach is complementary to other 
phenomenological investigations that have gone beyond classical
equilibrium thermodynamics.  Two particularly worth mentioning
theories are {\em steady state thermodynamics} \cite{oono} and the {\em finite time thermodynamics} \cite{ftt}.   The former has been shown to be consistent with stochastic processes with either
discrete or continuous state spaces \cite{ge_qian_10,sasa}; 
and the latter considered the notion of thermodynamic 
dissipation, especially in the coupling between a system and its 
baths, without actually requiring a time-dependent description.

	The stochastic diffusion theory presented in the present work
could provide a richer narrative for complex phenomenon which
has a stochastic dynamic description, but currently lacks a
concrete connection to vocabularies with mechanical and 
statistical thermodynamic implications, for example 
behavioral economics.  
Paraphrasing Montroll and Green \cite{montroll_green_54}: 
The aim of a stochastic thermodynamic theory is to develop a 
formalism from which one can deduce the collective 
behavior of complex systems composed of a large
number of individuals from a specification of the
component species, the laws of force which govern
interactions, and the nature of their surroundings.  Since the
work of Kirkwood \cite{kirkwood_35}, it has become 
clear that ``the laws of force'' can themselves emergent 
perperties with statistical (entropic) nature.  Indeed, 
Eq. \ref{1stlaw} could be recognized as one of such. 

Probability is a force of nature.  In the western legal system, 
this term refers to an event outside of human control for 
which no one can be held responsible.  Still, something 
no individual, or a small group of individuals, can be held 
responsible is nevertheless responsible by each and
every individual, together.  More is different.

\end{document}